\documentclass{PoS-hep}
\usepackage{amsmath}
\usepackage{bm}
\usepackage{epsfig}
\usepackage{graphics}
\usepackage{eufrak}
\usepackage{cite}
\usepackage{epsfig}
\usepackage{amssymb}
\usepackage{latexsym}
\usepackage{times}
\usepackage{slashed}
\usepackage{upgreek}
\usepackage{amsmath}
\usepackage{amsfonts}
\usepackage{amsbsy}
\usepackage{amscd}
\usepackage{bbm}

\newcommand{\eec}{\end{center}}
\newcommand{\bec}{\begin{center}}

\newcommand{\eem}{\end{matrix}}
\newcommand{\bem}{\begin{matrix}}
\newcommand{\eeq}{\end{equation}}
\newcommand{\beq}{\begin{equation}}
\newcommand{\ba}{\begin{array}}
\newcommand{\ea}{\end{array}}
\newcommand{\bea}{\begin{eqnarray}}
\newcommand{\eea}{\end{eqnarray}}
\newcommand{\baq}{\begin{eqnarray}}
\newcommand{\eaq}{\end{eqnarray}}
\newcommand{\beqs}{\begin{subequations}}
\newcommand{\eeqs}{\end{subequations}}
\newcommand{\bel}{\begin{align}}
\newcommand{\eal}{\end{align}}

\newcommand\eqs[2]{Eqs.~(\ref{#1}) and (\ref{#2})}

\newcommand{\ftn}{\footnotesize}
\newcommand{\nsz}{\normalsize}

\newcommand{\TeV}{{\mbox{\rm TeV}}}
\newcommand{\MeV}{{\mbox{\rm MeV}}}
\newcommand{\GeV}{{\mbox{\rm GeV}}}

\newcommand{\EeV}{{\mbox{\rm EeV}}}
\newcommand{\PeV}{{\mbox{\rm PeV}}}
\newcommand{\ZeV}{{\mbox{\rm ZeV}}}
\newcommand{\YeV}{{\mbox{\rm YeV}}}
\newcommand{\ReV}{{\mbox{\rm ReV}}}

\newcommand{\hz}{{\mbox{\rm Hz}}}
\newcommand{\sFref}[2]{Fig.~\ref{#1}-{\small \sf ({#2})}}

\newcommand{\etal}{{\it et al.\/}}

\def\lf{\left(}
\def\rg{\right)}
\newcommand\vev[1]{\langle {#1} \rangle}
\newcommand\vevi[1]{\langle {#1} \rangle_{\rm I}}
\newcommand{\Gr}{\ensuremath{\widetilde{G}}}

\newcommand{\ks}{\ensuremath{k_\star}}
\newcommand{\Gsm}{\ensuremath{\mathbb{G}_{\rm SM}}}

\newcommand{\Vhi}{\ensuremath{V_{\rm I}}}
\newcommand{\vf}{\ensuremath{V_{\rm F}}}
\newcommand{\Hhi}{\ensuremath{H_{\rm I}}}
\newcommand{\Whi}{\ensuremath{W_{\rm I}}}

\newcommand{\Vhio}{\ensuremath{V_{\rm I0}}}

\newcommand{\mP}{\ensuremath{m_{\rm P}}}

\newcommand{\Gu}{\ensuremath{\mathbb{G}_{\rm GUT}}}

\newcommand{\Glr}{\ensuremath{\mathbb{G}_{\rm LR}}}

\newcommand{\Glra}{\ensuremath{\mathbb{G}_{\rm L1R}}}

\newcommand{\suc}{\ensuremath{SU(3)_{\rm C}}}
\newcommand{\sur}{\ensuremath{SU(2)_{\rm R}}}

\newcommand{\ubl}{\ensuremath{U(1)_{B-L}}}
\newcommand{\ur}{\ensuremath{U(1)_{\rm R}}}
\newcommand{\vt}{\ensuremath{v_{\rm R}}}

\newcommand{\la}{\ensuremath{\lambda}}
\newcommand{\lm}{\ensuremath{\lambda_\mu}}
\newcommand{\aS}{\ensuremath{{\rm a}_S}}

\newcommand{\Gsn}{\ensuremath{\Gamma_{\rm I}}}
\newcommand{\msn}{\ensuremath{m_{\rm I}}}
\newcommand{\mgr}{\ensuremath{m_{3/2}}}
\newcommand{\mgri}{\ensuremath{m_{\rm I3/2}}}

\newcommand{\hd}{{\ensuremath{H_d}}}
\newcommand{\hu}{{\ensuremath{H_u}}}

\newcommand{\ns}{\ensuremath{n_{\rm s}}}
\newcommand{\as}{\ensuremath{\alpha_{\rm s}}}
\newcommand{\sni}{\ensuremath{\nu^c_i}}

\newcommand{\om}{\ensuremath{\omega}}

\newcommand{\br}{\ensuremath{{\sf B}_{\rm h}}}

\newcommand{\Trh}{\ensuremath{T_{\rm rh}}}

\newcommand{\sg}{\ensuremath{\sigma}}
\newcommand{\sgx}{\ensuremath{\sigma_\star}}
\newcommand{\sgc}{\ensuremath{\sigma_{\rm c}}}
\newcommand{\sgm}{\ensuremath{\sigma_{\rm max}}}

\newcommand{\ld}{\ensuremath{\lambda}}

\newcommand{\kp}{\ensuremath{\kappa}}

\newcommand{\mwr}{\ensuremath{{M_{W^\pm_{\rm R}}}}}
\def\ssb{\leavevmode\hbox{{\Large$\diagup$}{\kern-18pt\nsz SUSY}}}

\newcommand{\mss}{\ensuremath{\widetilde m}}

\newcommand{\hepph}[1]{{\ftn\tt hep-ph/#1}}

\newcommand{\arxiv}[1]{{\ftn\tt  arXiv:#1}}

\newcommand{\Eref}[1]{Eq.~(\ref{#1})}
\newcommand{\Sref}[1]{Sec.~\ref{#1}}
\newcommand{\Fref}[1]{Fig.~\ref{#1}}
\newcommand{\Tref}[1]{Table~\ref{#1}}
\newcommand{\cref}[1]{Ref.~\cite{#1}}
\newcommand{\crefs}[1]{Refs.~\cite{#1}}

\def\ths{{\theta_S}}

\def\Ka{K\"{a}hler potential}

\def\Kam{K\"{a}hler manifold}

\newcommand{\plk}{{\it Planck}}
\newcommand{\zm}{\ensuremath{Z_{-}}}

\newcommand{\lrh[1]}{\ensuremath{\lambda_{#1N^c}}}

\newcommand{\bdhh}{{\ensuremath{\normalsize I{\kern-2.9pt H}}}}

\newcommand{\phc}{\ensuremath{\Phi}}
\newcommand{\phcb}{\ensuremath{\bar\Phi}}
\newcommand{\what}{\ensuremath{\widehat}}

\newcommand{\mgro}{\ensuremath{m_{3/2}}}
\newcommand{\mz}{\ensuremath{m_{z}}}
\newcommand{\mzi}{\ensuremath{m_{{\rm I}z}}}
\newcommand{\mth}{\ensuremath{m_{\theta}}}
\newcommand{\mthi}{\ensuremath{m_{\rm I\theta}}}

\newcommand{\no}{\ensuremath{N}}

\def\al{{\alpha}}

\def\bz{{Z^*}}

\newcommand{\mcs}{\ensuremath{{\mu_{\rm cs}}}}
\newcommand{\gmcs}{\ensuremath{{G\mu_{\rm cs}}}}
\newcommand{\rcs}{\ensuremath{{r_{\rm cs}}}}
\newcommand{\ecs}{\ensuremath{{\epsilon_{\rm cs}}}}
\newcommand{\rms}{\ensuremath{{r_{\rm ms}}}}
\newcommand{\srms}{\ensuremath{{r_{\rm ms}^{1/2}}}}

\newcommand{\Ns}{\ensuremath{{N_{\rm I\star}}}}

\newcommand{\khh}{\ensuremath{K_{\rm H}}}

\newcommand{\dK}{\ensuremath{K_\mu}}

\newcommand{\dz}{\ensuremath{{\delta} z}}
\newcommand{\dzh}{\ensuremath{\what{\delta z}}}

\renewcommand{\Gsn}{\ensuremath{{\Gamma}_{\dz}}}

\newcommand{\hh}{{\ensuremath{
I{\kern-2.6pt h}}}}

\def\nano{{\sf\ftn NG15}}

\def\ligo{{\sf\ftn LVK}}
\newcommand{\ogw}{\ensuremath{\Omega_{\rm GW}h^2}}

\renewenvironment{subequations}{%
\refstepcounter{equation}%
\setcounter{parentequation}{\value{equation}}%
  \setcounter{equation}{0}
  \ignorespaces
}{%
  \setcounter{equation}{\value{parentequation}}%
  \ignorespacesafterend
}

\title{\boldmath \bfseries F-Term Hybrid Inflation, Metastable Cosmic Strings and Low
Reheating in View of ACT}

\ShortTitle{FHI, Metastable CSs \& low Reheating in View of
{\ftn\sffamily ACT}}

\author{\speaker{C. Pallis}\\
School of Civil Engineering, \\ Faculty of Engineering, \\
Aristotle University of Thessaloniki, \\ GR-541 24 Thessaloniki, GREECE \\
E-mail: \email{kpallis@auth.gr}}

\abstract{We consider the formation of metastable cosmic strings
in a left-right unified theory. The produced monopoles are diluted
by a stage of F-term hybrid inflation (FHI) which is realized
consistently with the SUSY breaking and a global $U(1)$ $R$
symmetry in the context of a $U(1)_{\rm R}\times U(1)_{B-L}$
extension of MSSM. The hidden sector \Kam\ enjoys an enhanced
$SU(1,1)/U(1)$ symmetry with the scalar curvature determined by
the achievement of a SUSY-breaking de Sitter vacuum without ugly
tuning. FHI turns out to be compatible with data -- including the
recent ACT results --, provided that the magnitude of the emergent
soft tadpole term is confined in the range $(0.1-70)$ TeV, and it
is accompanied with the production of cosmic strings. Their
dimensionless tension $\gmcs\simeq(1-11)\cdot10^{-8}$ interprets
the present observations from PTA experiments on the stochastic
background of gravitational waves. The $\mu$ parameter of MSSM
arises by appropriately adapting the Giudice-Masiero mechanism and
facilitates the out-of-equilibrium decay of the $R$ saxion at a
reheat temperature lower than about $34$ GeV. The SUSY mass scale
turns out to lie in the PeV region.
\\ \\{\sl\bfseries Published in}~~{\sl PoS  CORFU {\bf 2024}, 206 (2025)}
}

\FullConference{Corfu Summer Institute 2024 "School and Workshops
on Elementary Particle Physics and Gravity" (CORFU2024)
12 - 26 May and 25 August - 27 September, 2024\\
Corfu, Greece}

\begin{document}

\section{Introduction}

In this talk we attempt to bring together the recent data
\cite{pta, nano} on \emph{gravitational waves} ({\sf\ftn GWs}),
inflation \cite{plin, act} and \emph{Supersymmetry} ({\sf\ftn
SUSY}) breaking \cite{lhc}. The junction point is the metastable
\emph{cosmic strings} ({\sf\ftn CSs}) \cite{nano1} whose decay
interprets the recent data on GWs.  Their generation requires a
specific \emph{Grand Unified Theory} ({\sf\ftn  GUT})
construction -- see \Sref{intro1} -- in conjunction with a
suitable inflationary stage -- see \Sref{intro2}.


\subsection{PTA Data and Metastable CSs} \label{intro1}

The discovery of a background of GWs around the nanohertz
frequencies announced from several \emph{Pulsar Timing Array}
({\ftn \sf PTA}) experiments \cite{pta} -- most notably the
\emph{NANOGrav 15-years results} ({\ftn \sf NG15}) -- provides a
novel tool in exploring the structure of early universe. The
observations can be explained by gravitational radiation emitted
by topologically unstable superheavy CSs which may be formed
during the \emph{spontaneous symmetry breaking} ({\sf\ftn SSB})
chains of GUTs \cite{rachel} down to the \emph{Standard Model}
({\ftn\sf SM}) gauge group \Gsm. In particular, the observations
can be interpreted if the CSs are metastable. This type of CSs
arise from a SSB of the form
\beq \label{gmcs}\mathbb{G}\underset{\rm
MMs}{\xrightarrow{\hspace{0.2cm}\vev{{\rm
adj}(\mathbb{G})}\hspace{0.2cm}}} \mathbb{G}_{\rm int} \times
U(1)\underset{\rm CSs}{\xrightarrow{\hspace{0.4cm}\hspace{0.4cm}}}
\mathbb{G}_{\rm f}~~\mbox{with}~~\pi_1(\mathbb{G}/\mathbb{G}_{\rm
int})=\pi_1(\mathbb{G}/\mathbb{G}_{\rm f})=I\neq
\pi_1(\mathbb{G}_{\rm int} \times U(1)/\mathbb{G}_{\rm f}). \eeq
Here {\sf\ftn MMs} stands for \emph{Magnetic Monopoles} and ${\rm
adj}(\mathbb{G})$ for the adjoint representation of $\Gu$. Also
$\pi_n$ is the $n^{\rm th}$ homotopy group. {\sf\ftn MMs} are
produced at the first stage of the chain above since we expect
$$\pi_2(\mathbb{G}/\mathbb{G}_{\rm int} \times U(1))\neq I.$$ The
simplest way to implement such a SSB in a realistic particle model
is to identify
\beq \label{ulr}\mathbb{G}=\sur\times
\ubl~~\mbox{and}~~\mathbb{G}_{\rm int}\times U(1)=\ur \times
\ubl.\eeq
Here $\mathbb{G}$ may be embedded in the well-known Left-Right
gauge group -- cf. \cref{buch,nasri}
\beq \label{glr} \Glr:= SU(3)_{\rm C}\times SU(2)_{\rm L} \times
SU(2)_{\rm R} \times U(1)_{B-L} \eeq
and so  ${\rm adj}(\mathbb{G})$ may be identified by
$({\bf1,1,3},0)$. Although not cosmologically catastrophic, the
production of MMs may have an impact on the GWs if they are
inflated away. This is because they can appear on CSs via quantum
pair creation if the formatted CSs are metastable.

\subsection{CSs, Inflation and ACT
Data}\label{intro2}

A well-motivated inflationary model, which may dilute MMs and is
naturally followed by a GUT phase transition possibly leading to
the formation of CSs is \emph{F-term hybrid inflation} {\sf\ftn
FHI} \cite{hisusy} -- for reviews see \cref{hinova, hilaz}.
Therefore, we consider the implementation of FHI within the gauge
group
\beq\label{glra} \Glra:= SU(3)_{\rm C}\times SU(2)_{\rm L} \times
U(1)_{\rm R} \times U(1)_{B-L}\eeq
which results after the SSB of $\Glr$ from the \emph{vacuum
expectation value} ({\ftn\sf v.e.v}) $\vev{({\bf1,1,3},0)}:=\vt$.

For a reliable approach to FHI, soft SUSY-breaking terms and
\emph{Supergravity} ({\sf\ftn SUGRA}) corrections have to be taken
into account together with the \emph{radiative corrections} ({\ftn
\sf RCs}) employed in the original version of the model
\cite{hisusy}. Both former corrections above are of crucial
importance in order to reconcile the inflationary observables with
data \cite{sstad, mfhi, kaihi,shafiact} and are obviously related
to the adopted SUSY-breaking sector. Following \cref{asfhi, blfhi}
we here update -- in the light of \emph{Atacama Cosmology
Telescope} ({\sf\ftn ACT}) results \cite{act} -- the combination
of FHI and SUSY breaking using as a junction mechanism of the
(visible) \emph{inflationary sector} ({\ftn\sf IS}) and the
\emph{hidden sector} ({\sf\ftn HS}) a mildly violated $R$ symmetry
\cite{susyra} -- not to be confused with the gauge $\sur$ or $\ur$
symmetries included in $\Glr$ and $\Glra$, see \eqs{glr}{glra}
respectively.

Schematically, the steps of SSB in our setting can be demonstrated
as follows
\beq\begin{aligned}\Glr\times U(1)_{R} \times U(1)_{B}\times {\rm
SUGRA}~~&\underset{\rm MMs}{\xrightarrow{\hspace{0.75cm}
\vev{({\bf1,1,3},0)}=\vt\hspace{0.75cm}}}~~
\Glra\times U(1)_{R}\times U(1)_{B}\times {\rm SUGRA}\\
&-\hspace{-0.15cm}\lf\mbox{\rm\ftn
FHI}\rg\hspace{-0.15cm}\underset{\vev{Z},~ \mbox{\rm\ftn
CSs}}{{\xrightarrow{\hspace{0.4cm}\vev{\phcb}=\vev{\phc}=M
\hspace{0.4cm}}}}~~ \Gsm\times
U(1)_{B}\times ~~ \ssb\\
&~\underset{\rm RCs}{{\xrightarrow{\hspace{0.99cm}\vev{\hd},
\vev{\hu} \hspace{0.99cm}}}}~~\suc \times U(1)_{\rm EM}\times
U(1)_{B}. \label{chain1}\end{aligned}\eeq
The final transition to the present vacuum occurs via the
well-known radiative electroweak SSB -- {\sf\small EM} stands for
\emph{ElectroMagnetism}. Contrary to \cref{lrshafi, lrthi} we do
not specify the dynamics of the SSB of $\Glr$ but we incorporate
the mechanism of SUSY breaking as shown in the rightmost part of
the second line. Note that soft SUSY-breaking effects explicitly
break $U(1)_R$ to a discrete subgroup which is spontaneously
broken by the v.e.v of the sgoldstino, $Z$, too. Thanks to the
explicit violation of $U(1)_R$ in the \Ka\ -- see below -- no
problem with domain walls arises.

\subparagraph{}  We describe in \Sref{md} below our model. Then,
we analyze the inflationary -- see \Sref{fhi} -- and the
post-inflationary -- see \Sref{pres} -- stage of the cosmological
evolution within our setting. The spectrum of GWs obtained by the
CSs within our framework is presented in \Sref{css}. Our
conclusions are summarized in \Sref{con}.

\section{Particle Model}\label{md}

Here we determine the particle content, the superpotential, and
the \Ka\ of our model. These ingredients are presented in
Secs.~\ref{md1}, \ref{md2} and \ref{md3}.

\subsection{Particle Content}\label{md1}

As already mentioned, we focus on the second line of \Eref{chain1}
and therefore we consider a model invariant under the gauge group
$\Glra$ in \Eref{glra} which incorporates FHI. In addition to the
local symmetry, the model possesses the baryon and lepton number
symmetries and an $R$ symmetry $U(1)_{R}$. The representations
under $\Glra$ and the charges under the global symmetries of the
various matter and Higgs superfields of the model are presented in
Table~\ref{tab1}. Namely, the $i$th generation $SU(2)_{\rm L}$
doublet left-handed quark and lepton superfields are denoted by
$q_i$ and $l_i$ respectively, whereas the $SU(2)_{\rm L}$ singlet
antiquark [antilepton] superfields by $u^c_i$ and ${d_i}^c$
[$e^c_i$ and $\sni$] respectively. The electroweak Higgs
superfields which couple to the up [down] quark superfields are
denoted by $\hu$ [$\hd$]. Besides the MSSM particle content, the
model is augmented by seven superfields: a gauge singlet $S$,
three $\sni$'s, a pair of Higgs superfields $\phc$ and $\phcb$
which break $U(1)_{\rm R}\times U(1)_{B-L}$ and the goldstino
superfield $Z$. Note that the SM hypercharge $Q_Y$ is identified
as the linear combination $Q_Y=Q_{\rm R}+Q_{(B-L)}/2$ where
$Q_{\rm R}$ and $Q_{(B-L)}$ is the $\ur$ and $B-L$ charge
respectively.

\renewcommand{\arraystretch}{1.1}

\begin{table}[!t]
\begin{center}
\begin{tabular}{|c|c|c|c|c|}\hline
{\sc Super-}&{\sc Representations}&\multicolumn{3}{|c|}{\sc Global
Symmetries}\\\cline{3-5}
{\sc fields}&{\sc under $\Glra$}& {\hspace*{0.3cm}
$R$\hspace*{0.3cm} } &{\hspace*{0.3cm}$B$\hspace*{0.3cm}}&{$L$}
\\\hline\hline \multicolumn{5}{|c|}{\sc Matter
Superfields}\\\hline
{$l_i$} & {$({\bf 1, 2}, 0, -1)$} &$0$&{$0$}&{$1$}
\\
{$e^c_i$} &{$({\bf 1, 1}, 1/2, 1)$}& $0$&$0$ & $-1$ \\
{$\sni$} &{$({\bf 1, 1}, -1/2, 1)$}& $0$ &$0$ & $-1$\\
{$q_i$} & {$({\bf 3, 2}, 0 ,1/3)$} &$0$ &$1/3$&{$0$} \\
{$u^c_i$} &{$({\bf \bar 3, 1}, 1/2, -1/3)$}& $0$  &$-1/3$& $0$ \\
{$d^c_i$} &{$({\bf \bar 3, 1}, -1/2, -1/3)$}& $0$ &$-1/3$& $0$\\
\hline
\multicolumn{5}{|c|}{\sc Higgs Superfields}\\\hline
{$\hd$}&$({\bf 1, 2}, -1/2, 0)$& {$2$}&{$0$}&{$0$}\\
{$\hu$} &{$({\bf 1, 2}, 1/2, 0)$}& {$2$} & {$0$}&{$0$}\\
\hline
{$S$} & {$({\bf 1, 1}, 0, 0)$}&$2$ &$0$&$0$  \\
{$\Phi$} &{$({\bf 1, 1}, -1/2, 1)$}&{$0$} & {$0$}&{$-2$}\\
{$\bar \Phi$}&$({\bf 1, 1}, 1/2, -1)$&{$0$}&{$0$}&{$2$}\\\hline
\multicolumn{5}{|c|}{\sc Goldstino Superfield}\\\hline
{$Z$}&$({\bf 1, 1}, 0,0)$&{$2/\nu$}&{$0$}&{$0$}\\
\hline\end{tabular}
\end{center}
\caption[]{\sl \small The representations under $\Glra$ and the
extra global charges of the superfields of our model.}\label{tab1}
\end{table}
\renewcommand{\arraystretch}{1.}

\subsection{Superpotential}\label{md2}

The superpotential of our model respects totally the symmetries in
\Tref{tab1}. Most notably, it carries $R$ charge 2 and is linear
\emph{with respect to} ({\ftn\sf w.r.t.}) $S$ and $Z^{\nu}$. It
naturally splits into five parts:
\beq \label{Who} W=W_{\rm I} +W_{\rm H} +W_{\rm GH}+W_{\rm
MSSM}+W_{\rm MD},\eeq
where the subscripts ``I'' and ``H'' stand for IS and HS
respectively and the content of each term is specified as follows:

\subparagraph{\sf\ftn (a)} $\Whi$ includes the terms of the IS of
model \cite{hisusy}
\beqs\beq \Whi = \kp S\left(\bar
\Phi\Phi-M^2\right),\label{whi}\eeq
where $\kp$ and $M$ are free parameters which may be made positive
by field redefinitions.

\subparagraph{\sf\ftn (b)} $W_{\rm H}$ depends on the HS and is
written as \cite{susyra}
\beq W_{\rm H} = m\mP^2 (Z/\mP)^\nu. \label{wh} \eeq
Here $\mP=2.4~\ReV$ is the reduced Planck mass -- with
$\ReV=10^{18}~\GeV$ --, $m$ is a positive free parameter with mass
dimensions, and $\nu$ is an exponent which may, in principle,
acquire any real value if we consider $W_{\rm H}$ as an effective
superpotential valid close to the non-zero v.e.v of $Z$,
$\vev{Z}$.


\subparagraph{\sf\ftn  (c)} $W_{\rm GH}$ mixes the HS and the
gauge fields of the IS. It has the form
\beq W_{\rm GH} = -\la\mP(Z/\mP)^\nu \phcb\phc \label{wgh} \eeq
with $\la$ a real coupling constant.

\subparagraph{\sf\ftn (d)} $W_{\rm MSSM}$ contains the usual
trilinear terms of MSSM, i.e.,
\beq W_{\rm MSSM} = h_{ijD} {d}^c_i {q}_j \hd + h_{ijU} {u}^c_i
{q}_j \hu+h_{ijE} {e}^c_i {l}_j \hd. \label{wmssm}\eeq
The selected $R$ assignments in \Tref{tab1} prohibit the presence
in $W_{\rm MSSM}$ of the bilinear $\mu\hu\hd$ term of MSSM and
other unwanted mixing terms -- e.g. $\lm S\hu\hd$.

\subparagraph{\sf\ftn (e)} $W_{\rm MD}$ provides masses to
neutrinos, i.e.,
\beq W_{\rm MD} = h_{ij\nu} \sni l_j \hu+\lrh[i]\lf
S/\mP+(Z/\mP)^\nu\rg\lf\phcb
\nu^{c}_i\rg^2/\mP.\label{wmd}\eeq\eeqs
The first term in the right-hand side of \Eref{wmd} is responsible
for Dirac neutrino masses whereas the third one -- given that
$\vev{S}=0$ as we see in \Sref{des} below -- for the Majorana
masses -- cf. \crefs{mfhi, blfhi}. The scale of the latter masses
is intermediate since $\vev{\phcb}\sim1~\YeV$ and
$\vev{Z}\sim\mP$. The cooperation of both terms lead to the light
neutrino masses via the well-known (type I) seesaw mechanism.

\subsection{K\"{a}hler Potential}\label{md3}

The \Ka\ respects the $\Glra$, $B$ and $L$ symmetries in
\Tref{tab1} and breaks mildly the $R$ symmetry. It includes the
following contributions
\beq \label{Kho} K=K_{\rm I}+K_{\rm H}+\dK+|Y_\al|^2,\eeq
where the left-handed chiral superfields of MSSM are denoted by
$Y_\al$ with $\al=1,...,20$, i.e.,
\bea Y_\al= {l_i}, {e_i}^c, \nu_i^c, {q_i}, {d_i}^c, {u_i}^c,
\hd~\mbox{and}~\hu.\nonumber \eea
The individual contributions into $K$ can be specified as follows:

\subparagraph{\sf\ftn (a)} $K_{\rm I}$ depends on the fields
involved in IS -- cf. \Eref{whi}. We adopt the simplest possible
choice  -- cf. \cref{mfhi, hinova} -- which has the form
\beqs\beq K_{\rm I} = |S|^2+|\Phi|^2+|\bar\Phi|^2 \\
\label{ki} \eeq
and has zero contribution into the quadratic correction to the
inflationary potential -- see \Sref{fhi2} below.

\subparagraph{\sf\ftn (b)} $K_{\rm H}$ is devoted to HS. We adopt
the form introduced in \cref{susyra} where
\beq K_{\rm
H}=\no\mP^2\ln\lf1+\frac{|Z|^2-k^2\zm^4/\mP^2}{\no\mP^2}\rg
~~\mbox{with}~~Z_{\pm}=Z\pm Z^*. \label{khi} \eeq\eeqs
Here, $k>0$ mildly violates the $R$ symmetry endowing $R$ axion
with phenomenologically acceptable mass. The selected form of
$K_{\rm H}$ ensures -- as we see in \Sref{des} -- a de Sitter
vacuum of the whole field system with tunable cosmological
constant for
\beq
\no=\frac{4\nu^2}{3-4\nu}~~\mbox{with}~~\frac34<\nu<\frac32~~\mbox{for}~~\no<0.\label{no}
\eeq
Our favored $\nu$ range finally is $3/4<\nu<1$. Since $\no<0$,
$K_{\rm H}$ parameterizes the $SU(1,1)/U(1)$ hyperbolic \Kam\ for
$k\sim0$.

\subparagraph{\sf\ftn (c)} $\dK$ includes higher order terms which
generate the needed mixing term between $\hu$ and $\hd$ in the
lagrangian of MSSM \cite{susyra} and has the form
\beq \dK=\lm\lf\bz/\mP\rg^{2\nu}\hu\hd\ +\ {\rm
h.c.},\label{dK}\eeq
where the dimensionless constant $\lm$ is taken real for
simplicity.

\subparagraph{} The total $K$ in \Eref{Kho} enjoys an enhanced
symmetry for the $\bar Y^A$ and $Z$ fields where
$A=\al,\phcb,\phc, S$. Namely, this symmetry can be written as
\beq  \prod_A U(1)_{\bar Y^A}\times  \lf SU(1,1)/U(1)\rg_Z,\eeq
where the indices indicate the moduli which parameterize the
corresponding manifolds. Thanks to this symmetry, mixing terms
allowed by the $R$ symmetry can be ignored.

\section{Inflation Analysis}\label{fhi}

It is well known \cite{hisusy} that in global SUSY FHI takes place
for sufficiently large $|S|$ values along a F- and D- flat
direction of the SUSY potential
\begin{equation} \label{v0}\bar\Phi={\Phi}=0,~~\mbox{where}~~ V_{\rm SUSY}\lf{\Phi}=0\rg:= V_{\rm I0}=\kp^2
M^4~~\mbox{and}~~\Hhi=\sqrt{\Vhio/3\mP^2}\eeq
are the constant potential energy density and correspoding Hubble
parameter which drive FHI -- the subscript $0$ means that this is
the tree level value. In the present context, though, prominent
role in the investigation plays the F-term SUGRA potential $\vf$
which is minimized under the conditions displayed in
Sec.~\ref{fhi1} and then in \Sref{fhi2} we give the final form of
the inflationary potential. Lastly, we present our results in
\Sref{fhi4} imposing a number of constraints listed in
\Sref{fhi3}.

\subsection{Hidden Sector's Stabilization}\label{fhi1}

If we analyze $S$ and $Z$ as follows
\beq S=\sg\
e^{i\ths/\mP}/\sqrt{2}~~\mbox{and}~~Z=(z+i\theta)/\sqrt{2},\label{para}\eeq
we can show \cite{asfhi} that $\vf$ is minimized  along the
inflationary trajectory in \Eref{v0} for
\beq \label{veviz}
\vevi{\ths}=\vevi{\theta}=0~~\mbox{and}~~\vevi{z}\simeq\lf\sqrt{3}\cdot2^{\nu/2-1}\Hhi/m\nu\sqrt{1-\nu}\rg^{1/(\nu-2)}\mP.\eeq
Note that $\nu<1$ assures a real value of $\vevi{z}$ with
$\vevi{z}\ll\mP$ since $\Hhi/m\ll1$.

The (canonically normalized) components of sgoldstino, acquire
masses squared, respectively,
\beqs\beq\mzi^2\simeq6(2-\nu)\Hhi^2~~\mbox{and}~~
\mthi^2\simeq3\Hhi^2-
m^2\lf8\nu^2\mP^2-3\vevi{z}^2\rg\frac{4\nu(1-\nu)\mP^2+(1-96k^2\nu)\vevi{z}^2}{2^{3+\nu}\nu\mP^{2\nu}\vevi{z}^{2(2-\nu)}},
\label{mz8i} \eeq
whereas the mass of $\Gr$ turns out to be
\beq \mgri\simeq \lf
\nu(1-\nu)^{1/2}m^{2/\nu}/\sqrt{3}\Hhi\rg^{\nu/(2-\nu)}.\eeq\eeqs
For the benchmark point presented in \Tref{tab} we can numerically
verify that $\mzi\gg\Hhi$ and $m_{\rm I\theta}\simeq\Hhi$ given
that $\Hhi=1.05~\EeV$ consistently with the inflationary
requirements in \Sref{fhi2}. Therefore, isocurvature perturbations
do not constrain the parameters \cite{asfhi}.

\renewcommand{\arraystretch}{1.1}
\begin{table} \bec \begin{tabular}{|c|c|c|c|}\hline
\multicolumn{4}{|c|}{\sc Model Parameters} \\ \hline\hline
$\ld/10^{-12}$&$M/\YeV$&$m/\PeV$&$\as/\TeV$\\\hline
$1.4$&$2.3$&$3.5$&$25.6$\\\hline\hline
\multicolumn{4}{|c|}{\sc HS Parameters During FHI} \\ \hline\hline
$\vevi{z}/10^{-3}\mP$&$\mzi/\EeV$&$\mthi/\EeV$&$\mgri/\EeV$\\\hline
$2$&$2.7$&$0.5$&$11.2$\\\hline\hline
\multicolumn{4}{|c|}{\sc Quantities Related to FHI}\\\hline\hline
$\sgx/\sqrt{2}M$&$\as/10^{-4}$&$r/10^{-11}$&$\gmcs/10^{-8}$\\\hline
$1.097$&$-2.3$&$2.5$&$7.4$\\\hline\hline
\multicolumn{4}{|c|}{\sc Spectrum at the Vacuum}\\\hline\hline
$\msn/\ZeV$&$\mz/\PeV$&$\mth/\PeV$&$\mgr/\PeV$\\\hline
$5.2$&$6.2$&$8.8$&$5.6$\\\hline\hline
\multicolumn{4}{|c|}{\sc Reheat Temperature $\Trh/\GeV$
}\\\hline\hline
\multicolumn{2}{|c|}{For $\mu/\mss=3$}&\multicolumn{2}{|c|}{For
$\mu/\mss=1/3$}\\\hline
\multicolumn{2}{|c|}{$3$}&\multicolumn{2}{|c|}{$0.4$}\\\hline
\end{tabular}\eec
\caption{\sl Parameters and quantities of interest for our model
with fixed $\kp=0.001$, $\nu=7/8$ and $k=0.1$ resulting to
$\ns=0.974$ for $\Ns\simeq41$ -- recall that
$1~\YeV=10^3~\ZeV=10^6~\EeV=10^9~\PeV=10^{15}~\GeV$.}\label{tab}
\end{table}
\renewcommand{\arraystretch}{1.}

\subsection{Inflationary Potential}\label{fhi2}

Expanding $\vf$ for low $S$ values, introducing the canonically
normalized inflaton $\sg=\sqrt{2}|S|$ and taking into account the
RCs \cite{hisusy} we derive \cite{asfhi} the inflationary
potential $V_{\rm I}$ which can be cast in the form
\beq\label{vol} V_{\rm I}\simeq V_{\rm I0}\left(1+C_{\rm
RC}+C_{\rm SSB}+C_{\rm SUGRA}\right),~~\mbox{where}\eeq

\subparagraph{\sf\ftn (a)} $C_{\rm RC}$ includes the RCs which may
be written as \cite{hisusy, hilaz,hinova}
\beqs\beq \label{crc}C_{\rm RC}=
{\kappa^2\over 128\pi^2}\lf8\ln{\kp^2 M^2\over
Q^2}+8x^2\tanh^{-1}\lf\frac{2}{x^2}\rg-4(\ln4-x^4\ln
x)+(4+x^4)\ln(x^4-4)\rg\eeq
with $x=\sigma/ M>\sqrt{2}$. For $x\leq\sqrt{2}$, the path in
\Eref{v0} develops a tachyonic instability occurs in the mass
spectrum of the $\phcb-\phc$ system triggering thereby the
$\ur\times\ubl$ phase transition which generates the desired CSs.

\subparagraph{\sf\ftn (b)} $C_{\rm SSB}$ comes from the soft
SUSY-breaking effects \cite{sstad} and can be parameterized as
follows:
\beq \label{cssb}C_{\rm SSB}=
m_{\rm I3/2}^2 {\sg^2\over 2V_{\rm I0}}-{\rm a}_S {\sigma
\over\sqrt{2V_{\rm I0}}}~~\mbox{with}~~{\rm
a}_S=2^{1-\nu/2}m\frac{\vevi{z}^\nu}{\mP^\nu}\lf1+\frac{\vevi{z}^2}{2N\mP^2}\rg
\lf2-\nu-\frac{3\vevi{z}^2}{8\nu\mP^2}\rg\eeq
the tadpole parameter. In the expression above, we take care to
minimize $\Vhi$ w.r.t $\theta_S$ in \Eref{para} which is assumed
to be constant during FHI -- cf. \cref{kaihi}.

\subparagraph{\sf\ftn (c)} $C_{\rm SUGRA}$ is the remaining --
i.e., after subtracting $C_{\rm SSB}$ -- SUGRA correction, which
is
\beq \label{csugra} C_{\rm
SUGRA}=c_{2\nu}\frac{\sg^2}{2\mP^2}+c_{4\nu}\frac{\sg^4}{4\mP^4}~~\mbox{with}~~
 c_{2\nu}=\frac{\vevi{z}^2}{2\mP^2}~~\mbox{and}~~c_{4\nu}=\frac12\lf1+\frac{\vevi{z}^2}{\mP^2}\rg.
\eeq\eeqs
FHI is feasible thanks to the stabilization of $z$ in \Eref{veviz}
to low enough values and the minimality of $K_{\rm I}$ in
\Eref{ki} which offers the magic cancellation \cite{hinova} of the
quadratic correction in $C_{\rm SUGRA}$.

\subsection{Observational Requirements}\label{fhi3}

The inflationary part of our setting can be constrained by
imposing the following observational requirements:

\subparagraph{\sf\ftn (a)} The number of e-foldings elapsed from
horizon exit of $\ks=0.05/{\rm Mpc}$ $k$ until the end of
inflation at $\sigma_{\rm f}\simeq\sgc$ have to be enough to
resolve the problems of the Standard Big Bang, i.e., \cite{plin}:
\begin{equation}  \label{Nhi}
\Ns=-\int_{\sigma_{\star}}^{\sigma_{\rm f}}
\frac{d\sigma}{m^2_{\rm P}}\: \frac{V_{\rm I}}{V'_{\rm
I}}\simeq41+{2\over 3}\ln{V^{1/4}_{\rm I0}\over{0.1~{\rm YeV}}}+
{1\over3}\ln {T_{\rm rh}\over{1~{\rm GeV}}},
\end{equation}
where the prime denotes derivation w.r.t. $\sigma$ and  $\sgx$ is
the value of $\sigma$ when $\ks$ crosses outside the horizon of
FHI. Here, we have assumed that the ``reheating temperature'' due
to the decay of IS after FHI  is lower than $\rho_{z\rm I}$-- see
\Eref{rhoz} below -- and, thus, we obtain just matter domination
after the end of FHI and before the complete decay of $\dzh$ which
yields the correct reheating temperature $T_{\rm rh}\sim1~\GeV$.

\subparagraph{\sf\ftn (b)} The amplitude $A_{\rm s}$ of the power
spectrum of the curvature perturbation generated by $\sigma$
during FHI must be appropriately normalized \cite{act} , i.e.,
\begin{equation} \label{Prob}
\sqrt{A_{\rm s}}= \frac{1}{2\sqrt{3}\pi m^3_{\rm P}}\;
\left.\frac{V_{\rm I}^{3/2}(\sigma_\star)}{|V'_{\rm
I}(\sigma_\star)|}\right.\simeq\: 4.618\cdot 10^{-5}.
\end{equation}

\subparagraph{\sf\ftn (c)} The remaining observables -- the scalar
spectral index $\ns$, its running $\as$, and the scalar-to-tensor
ratio $r$ -- which are calculated by the following standard
formulas
\beq \label{ns}  \ns=1-6\epsilon_\star\ +\ 2\eta_\star,
\as={2}\left(4\eta_\star^2-(\ns-1)^2\right)/3-2\xi_\star~~\mbox{and}~~
r=16\epsilon_\star\eeq
(where $\xi\simeq m_{\rm P}^4~V'_{\rm I} V'''_{\rm I}/V^2_{\rm I}$
and all the variables with the subscript $\star$ are evaluated at
$\sigma=\sgx$) must be in agreement with the combination of \plk,
ACT, DESI and {\sc Bicep}/{\it Keck} telescopes ({\sf\ftn
P-ACT-LB-BK18}) data, i.e., \cite{act}
\beq \label{nswmap} \ns=0.974\pm0.0068~~\mbox{and}~~r\lesssim0.038
\eeq at 95$\%$ \emph{confidence level} ({\sf\ftn c.l.}) with
negligible $|\as|\ll0.01$.

\subsection{Results}\label{fhi4}

The inflationary part of our model depends on the parameters --
cf. \cref{mfhi} -- $\kappa,~M~~\mbox{and}~~\aS$, where $\aS$ can
be derived from $m$ and $\nu$ via the rightmost relation in
\Eref{cssb}. Enforcing Eqs.~(\ref{Nhi}) and (\ref{Prob}) we can
restrict $M$ and $\sgx$ as functions of $\kappa$ and $\aS$. On the
other hand, the correct $\ns$ values in \Eref{nswmap} are attained
if FHI becomes of hilltop type \cite{mfhi, kaihi} via a careful
selection of $\aS$. I.e., $\Vhi$ is non-monotonic and develops a
maximum at $\sgm$ and a minimum at $\sg_{\rm min}\gg\sgm$. For
$\sg>\sg_{\rm min}$, $\Vhi$ becomes a monotonically increasing
function of $\sg$ and so its boundedness is assured. FHI takes
place for $\sg<\sgm$. The achievement of successful hilltop FHI
requires the establishment of the hierarchy $\sgc<\sgx<\sgm$. As a
consequence, the obtained $\ns$ values decline from the prediction
$\ns\simeq1-1/\Ns$ of the original model \cite{hisusy,shafiact}
which provides the correct value in \Eref{nswmap}, though, for the
$\Ns\sim41$.

\begin{figure}[!t]\vspace*{-.18in}
\hspace*{-.12in}
\begin{minipage}{8in}
\epsfig{file=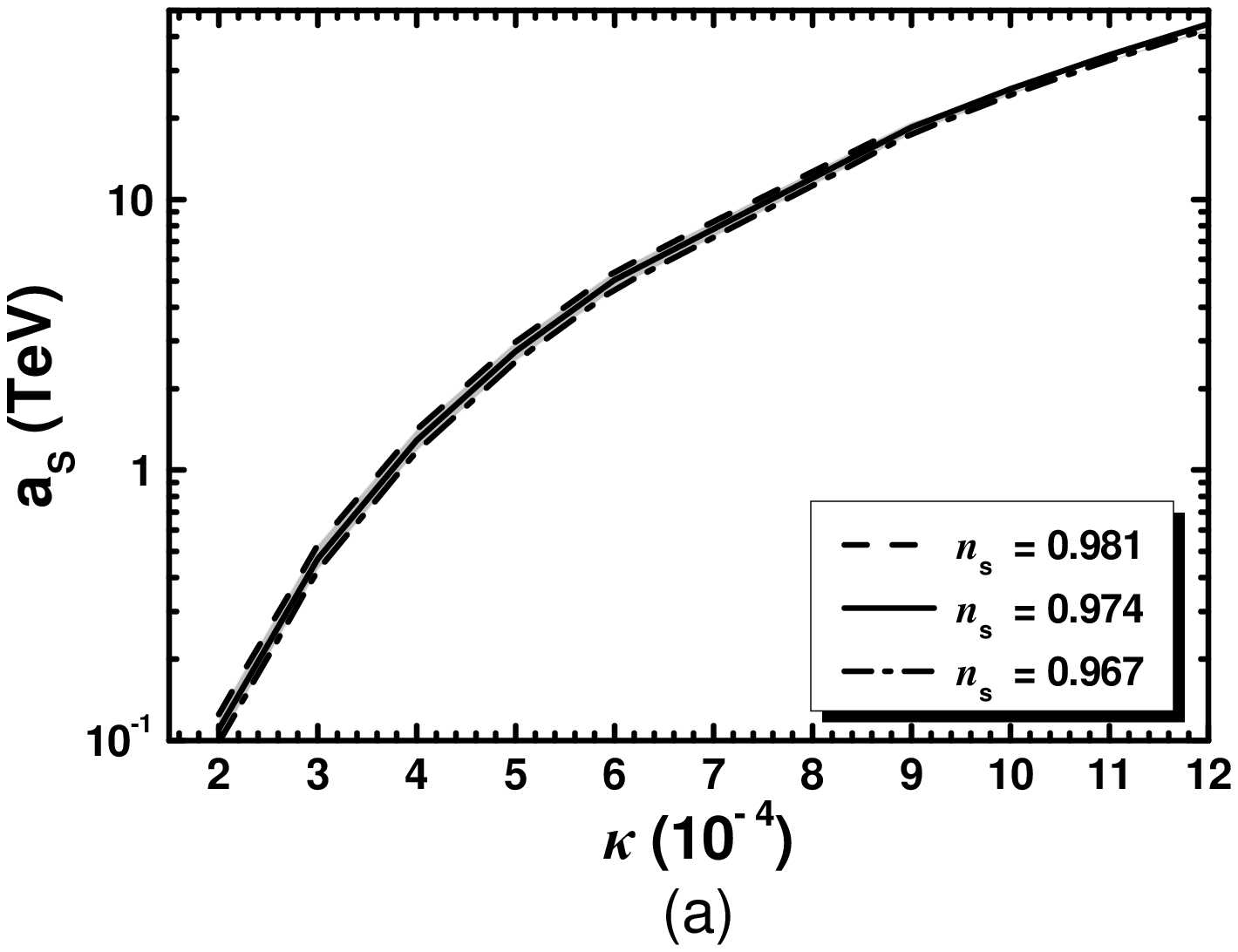,height=3.6in,angle=-90}
\hspace*{-1.3cm}
\epsfig{file=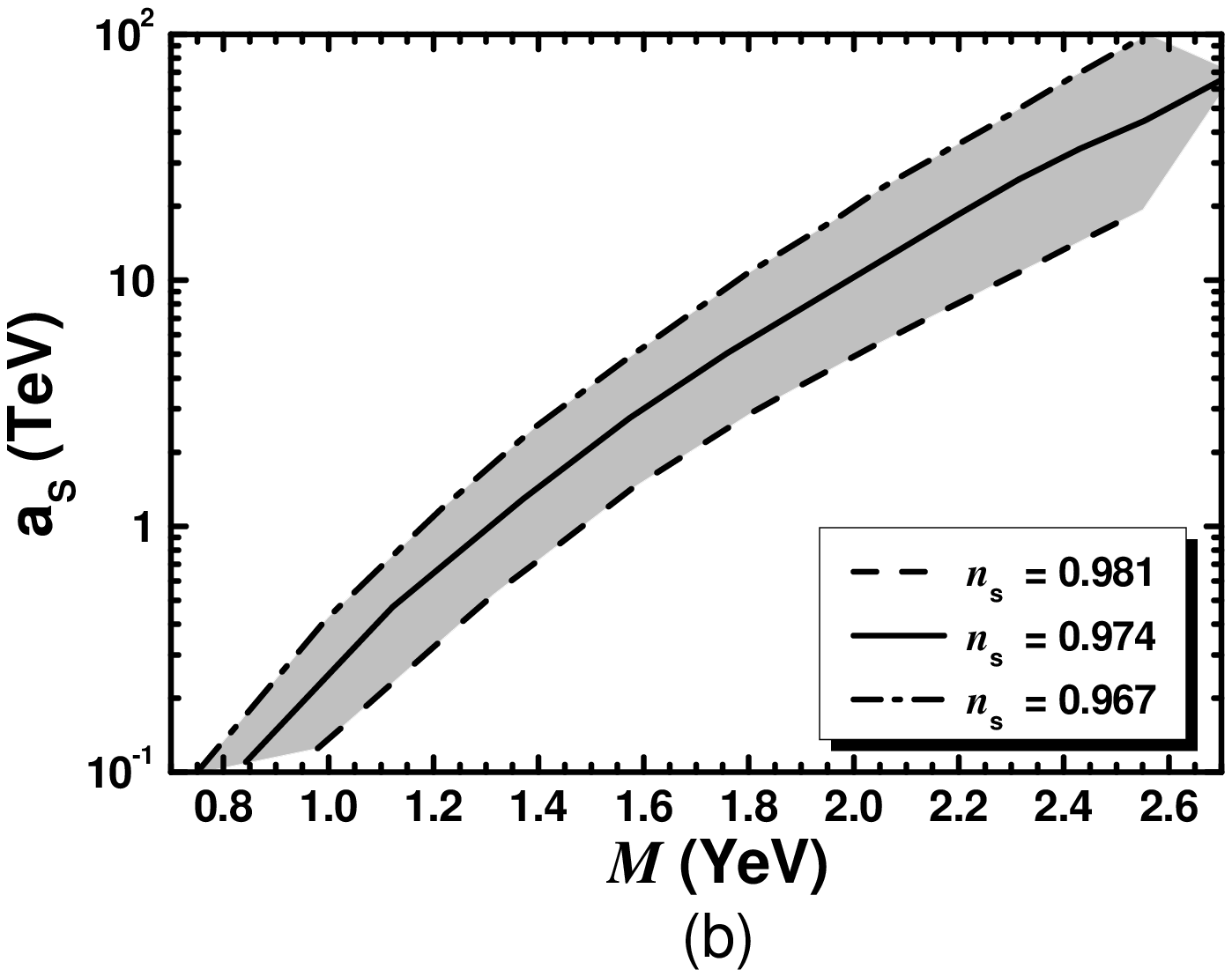,height=3.6in,angle=-90} \hfill
\end{minipage}
\hfill \caption{\sl\small Allowed (shaded) regions as determined
by Eqs.~(3.7), (3.8) and (3.10) in the $\kp-\aS$ {\sffamily\ftn
(a)} and $M-\aS$ {\sffamily\ftn (b)}. The conventions adopted for
the various lines are also shown. }\label{fig1}
\end{figure}

The parameter space of our model allowed by the inflationary
requirements in Eqs.~(\ref{Nhi}), (\ref{Prob}) and (\ref{nswmap})
is delineated in \Fref{fig1} where we present the gray shaded
regions in the $\kappa-\aS$ and $M-\aS$ plane  -- see
\sFref{fig1}{a} and \sFref{fig1}{b} respectively.  The boundaries
of the allowed areas are determined by the dashed [dot-dashed]
line corresponding to the upper [lower] bound on $n_{\rm s}$ in
Eq.~(\ref{nswmap}). We also display by solid lines the allowed
contours for the central value of $\ns$ in \Eref{nswmap}. The
maximal $r$'s are encountered in the upper right end of the dashed
lines -- corresponding to $\ns=0.981$. On the other hand, the
maximal $|\as|$'s are achieved along the dot-dashed lines and the
minimal value is $\as=-3.2\cdot10^{-4}$. Summarizing our findings
from Fig.~\ref{fig1} for central $\ns$ in \Eref{nswmap} we end up
with the following ranges:
\beq\label{res} 0.84\lesssim {M/{\rm YeV}}\lesssim2.72,~~2\lesssim
{\kp/10^{-4}}\lesssim14~~\mbox{and}~~0.11\lesssim{\aS/\TeV}\lesssim70.\eeq
The lower bounds of these inequalities  are expected to be
displaced to slightly larger values due to the constraints in
\Eref{tns} -- see \Sref{reh} below -- which are not considered
here for the shake of generality.

\section{Post-Inflationary Era}\label{pres}

The coupling of HS to IS not only explains the appearance of the
terms $C_{\rm SSB}$ and $C_{\rm SUGRA}$ in $\Vhi$ but also has
far-reaching consequences regarding the problem of DE, the
reheating in our scenario and the SUSY-mass scale. These issues
are analyzed in \Sref{des}, \ref{reh} and \ref{susy} below.

\subsection{\boldmath SUSY and $\Glra$ Breaking}\label{des}

The minimization of the F--term (tree level) SUGRA scalar
potential $V_{\rm F}$ \cite{asfhi}, which includes contributions
from both the HS and the IS, yields the vacuum of our model which
is determined by the conditions
\beq
\label{vevsg}\left|\vev{\Phi}\right|=\left|\vev{\bar\Phi}\right|=M,~~
\vev{\sg}\simeq2^{(1-\nu)/2}\lf m-\ld\mP\rg\
z^\nu/\mP^{\nu}~~\mbox{and}~~\vev{z}=2\sqrt{2/3}|\nu|\mP\eeq
with $\vev{\theta}=\vev{\theta_S}=0$. Contrary to the status in
the absence of IS \cite{susyra}, the vacuum potential energy
density here is not zero but constant with value
\beq \vev{\vf}=\lf\frac{16\nu^{4}}{9}\rg^\nu \lf\frac{\ld
M^2-m\mP}{\kp\mP^2}\rg^2\om^N\mP^2\lf\ld\mP-m\rg^2~~
\mbox{with}~~\om\simeq\frac{2(3-2\nu)}{3}.\label{vcc}\eeq
Tuning $\ld$ to a value $\ld\sim m/\mP\simeq10^{-12}$ we may wish
identify $\vev{\vf}$ with the DE energy density \cite{act}
\beq \label{omde} \vev{\vf}\simeq10^{-120}\mP^4.\eeq
Therefore, we obtain a dS vacuum with $\vev{\sg}\simeq0$ which
resolves without extensive tuning the notorious DE problem. At
this vacuum we can estimate the masses of the gravitino, the
sgoldstino (or $R$ saxion), the pseudo-sgoldstino (or $R$ axion)
and the IS which are respectively \cite{susyra}
\beqs\bea \label{mgr} \mgro&\simeq& 2^{\nu}3^{-\nu/2}
|\nu|^{\nu}m\omega^{N/2},~\mz\simeq\frac{3\om}{2\nu}\mgro,
~\mth\simeq12k\om^{3/2}\mgro\\
\mbox{and}~~ \msn&=&e^{{\khh}/{2\mP^2}}\sqrt{2}\lf\kp^2
M^2+(4\nu^{2}/3)^\nu(1+4M^2/\mP^2)m^2\rg^{1/2}.
 \label{mI} \eea\eeqs
Typical values for the masses above are given in \Tref{tab}, where
we see that besides $\msn$, which is of order $1~\ZeV$, the other
masses are of order $1~\PeV$.

At the vacuum of \Eref{vevsg} $\dK$ in \Eref{dK} gives rise to a
non-vanishing $\mu$ term in the superpotential whereas the
contributions $W_{\rm MSSM}$ and $|Y_\al|^2$ of $W$ and $K$ in
\eqs{Who}{Kho} lead to a common soft SUSY-breaking mass parameter
$\mss$. Namely, we obtain \cite{susyra,asfhi}
\beq W\ni \mu  H_u H_d ~~\mbox{with}~~|\mu|=
\lm\lf{4\nu^2}/{3}\rg^\nu(5-4\nu)\mgr~~\mbox{and}~~
\mss=\mgr.\label{mssi}\eeq
The latter quantity indicatively represents the mass level of the
SUSY partners and allows us to connect the inflationary model with
the SUSY phenomenology via \eqs{cssb}{res} -- see below.

\subsection{Low Reheating}\label{reh}

Soon after FHI, the IS and $\dzh$ enter into an oscillatory phase
about their minima in \Eref{vevsg} and eventually decay. Due to
the large $\vev{z}\sim\mP$ in \Eref{vevsg}, the energy density of
the $z$ condensate at the onset of oscillations, $\rho_{z\rm I}$,
-- when $H_{z\rm I}\sim\mz$ -- dominates the corresponding
universal energy density $\rho_{z\rm It}$. Indeed,
\beq \label{rhoz} \rho_{z\rm
I}\sim\mz^2\vev{z}^2\sim\mz^2\mP^2~~\mbox{and}~~\rho_{z\rm
It}=3\mP^2H_{z\rm I}^2\simeq3\mP^2\mz^2.\eeq
Due to weakness of the $z$ interactions, the reheating temperature
which is determined by the decay width of $\dzh$, $\Gsn$, is quite
suppressed -- see values of \Tref{tab} for two extremal
$\mu/\mgro$ values. Indeed,
\beq \label{Trh} \Trh= \left({72/5\pi^2g_{\rm
rh*}}\right)^{1/4}\Gsn^{1/2}\mP^{1/2},~~\mbox{with
\cite{asfhi}}~~\Gsn\sim\lm^2\mz^3/\mP^2 ~~\mbox{and}~~g_{\rm
rh*}\simeq10.75-100.\eeq
Here $\Gsn$ is the decay width of $\dzh$ -- cf. \cref{baerh, nsrh}
-- and $g_{\rm rh*}$ counts the effective number of the
relativistic degrees at $\Trh$. This fact jeopardizes the
successes of the standard \emph{Big Bang Nucleosynthesis}
({\sf\ftn BBN}) which requires \cite{nsref}
\beq \Trh\geq4.1~\MeV~~\mbox{for}~~\br=1~~\mbox{and}~~
\Trh\geq2.1~\MeV~~\mbox{for}~~\br=10^{-3},\label{tns}\eeq
where $\br$ is the hadronic branching ratio and large
$\mz\sim0.1~\PeV$ is assumed. Moreover, in order to protect our
setting from the so-called \cite{koichi} moduli-induced $\Gr$
problem, we kinematically block the decay of $\dzh$ into $\Gr$'s
selecting $\nu>3/4$ which ensures $\mz<2\mgr$ -- see \Eref{mgr}.

Taking $\kp$ and $\aS$ -- or $\mz$ values via \eqs{cssb}{mgr} --
allowed by the inflationary part of our model we find the allowed
region in $\kp -\Trh$ plane, displayed in \sFref{fig2}{a}, for
$\nu=7/8$. The boundary curves of the allowed region correspond to
$\mu=\mss/3$ or $\lm=0.22$ (dot-dashed line) and $\mu=3\mss$ or
$\lm=1.96$ (dashed line). We see that there is an ample parameter
space consistent with the BBN bounds depicted by two horizontal
lines depending on the values of $\br$ in \Eref{tns}. The maximal
value of $\Trh$ for the selected $\nu$ is obtained for $\mu=3\mss$
and is estimated to be $T_{\rm rh}^{\max}\simeq11~\GeV$.
Obviously, reducing $\mu$ below $\mss/3$, the parameters $\lm$,
$\Gsn$ and so $\Trh$ decrease too and the slice cut by the BBN
bounds in \Eref{tns} increases.

\begin{figure}[!t]\vspace*{-.18in}
\hspace*{-.12in}
\begin{minipage}{8in}
\epsfig{file=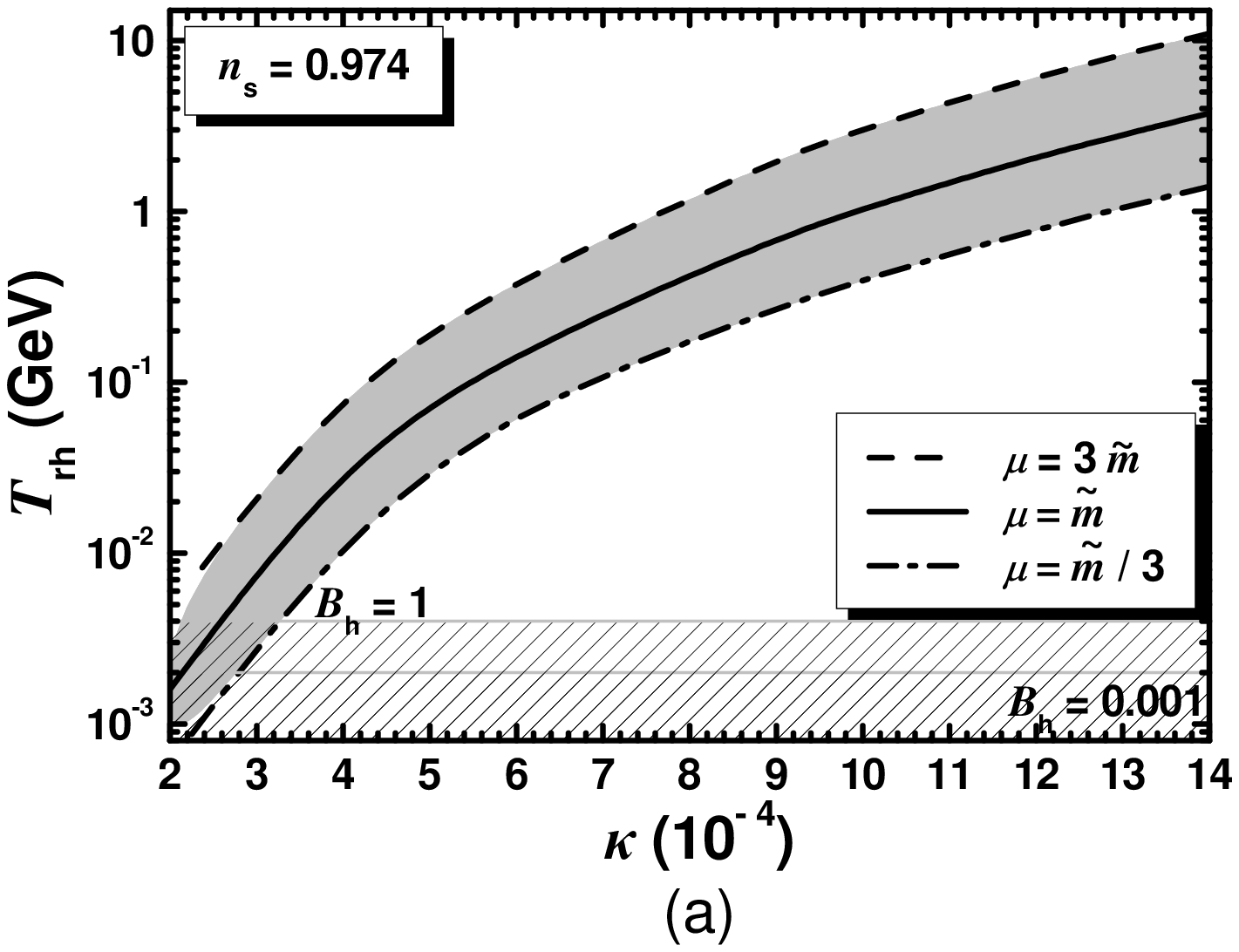,height=3.58in,angle=-90}
\hspace*{-1.3cm}
\epsfig{file=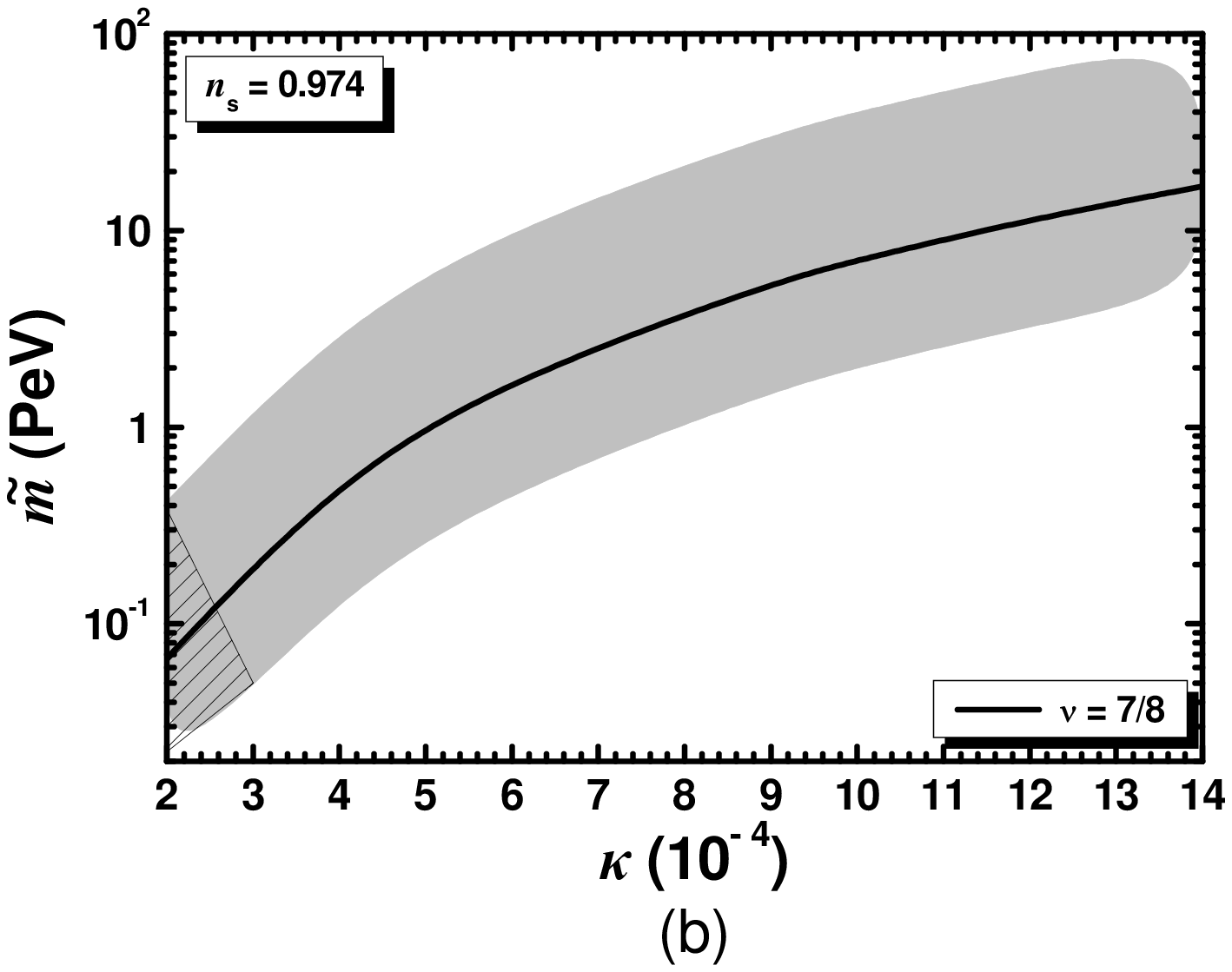,height=3.58in,angle=-90} \hfill
\end{minipage}
\hfill \caption{\sl\small Allowed (shaded) regions in the
$\kp-\Trh$ {\sffamily\ftn (a)} and $\kp-\mss$ {\sffamily\ftn (b)}
plane compatible with the inflationary requirements in Sec.~3.3
for $\ns=0.974$. We take {\sffamily\ftn (a)} $\nu=7/8$ and
$\mu=\mss$ (solid line), $\mu=\mss/3$ (dot-dashed line) or
$\mu=3\mss$ (dashed line) whereas the BBN lower bounds on $\Trh$
for hadronic branching ratios $\br=1$ and $0.001$ are also
depicted by two thin lines {\sffamily\ftn (b)} $\mss/3\leq
\mu\leq3\mss$ and $3/4<\nu<1$ whereas the allowed contour for
$\nu=7/8$ is depicted by a solid line and the region excluded by
BBN for $\br=0.001$ is hatched.}\label{fig2}
\end{figure}

\subsection{SUSY-Mass Scale}\label{susy}

As mentioned above, the restriction of the $\aS$ values in
Fig.~\ref{fig1} -- see also \Eref{res} -- via the attaintment of
successful FHI and the direct interconnection of $\aS$ with $m$
and $\mss$ via \eqs{cssb}{mssi} give us the opportunity to gain
information about the mass scale of SUSY particles. Indeed, taking
into account \Eref{veviz}, which results to
$\vevi{z}/\mP\sim10^{-3}$, and \Eref{res} we expect that
$m\sim1~\PeV$ and so $\mss\sim1~\PeV$ as well. This expectation is
verified numerically in \sFref{fig2}{b}, where we delineate the
gray-shaded region allowed by the inflationary requirements of
\Sref{fhi3} for $\ns=0.974$ by varying $\nu$ and $\mu$ within
their possible respective margins
\beq \label{susy1}
0.75<\nu<1~~\mbox{and}~~1/3\lesssim\mu/\mss\lesssim3. \eeq
Obviously the lower boundary curve of the displayed region is
obtained for $\nu\simeq0.751$, whereas the upper one corresponds
to $\nu\simeq0.99$. The hatched region is ruled out by \Eref{tns}.
All in all, we obtain the predictions
\beq0.16\lesssim\mss/\PeV\lesssim 9.2~~\mbox{with}~~T_{\rm
rh}^{\max}\simeq34.\label{susyres}\eeq
These results on $\mss$ in conjunction with the necessity for
$\mu\sim\mss$, established in \Sref{reh}, hint towards the
\PeV-scale MSSM. Our findings are compatible with the mass of the
Higgs boson discovered in LHC \cite{strumia} for degenerate
sparticle spectrum, $1\leq\tan\beta\leq50$ and variable stop
mixing.

\section{GWs from Metastable CSs}\label{css}

The $U(1)_{\rm R}\times U(1)_{B-L}$ breaking which occurs for
$\sg\simeq\sgc$ produces a network of CSs which can be considered
as metastable due to the embedding of $\Glra$ in $\Glr$ -- see
\Eref{chain1}. The dimensionless tension $\gmcs$ of the CSs
produced at the end of FHI can be estimated by \cite{mfhi}
\begin{equation} \label{mucs} \gmcs \simeq
\frac12\lf\frac{M}{\mP}\rg^2\ecs(\rcs)~~\mbox{with}~~\ecs(\rcs)=\frac{2.4}{\ln(2/\rcs)}~~
\mbox{and}~~\rcs=\kappa^2/2g^2\leq10^{-2},\end{equation}
where we take into account that $(B-L)(\phc)=1$ in accordance with
the coupling in \Eref{wmd}. Also $G=1/8\pi\mP^2$ is the Newton
gravitational constant and $g\simeq0.7$ is the gauge coupling
constant at a scale close to $M$. For the parameters in \Eref{res}
we find
\beq 0.81\lesssim\gmcs/10^{-8}\lesssim11.\label{rescs}\eeq

On the other hand, an explanation \cite{nano1} of the recent
\nano\  \cite{nano, pta} on stochastic GWs entails
\beq  10^{-8}\lesssim  \gmcs\lesssim 2.4\cdot
10^{-4}~~\mbox{for}~~8.2\gtrsim\sqrt{\rms}\gtrsim7.5~~ \mbox{at
95$\%$ c.l.},\label{kai} \eeq
where the metastability factor $\rms$ is calculated within our
scheme via the relation
\begin{equation} \label{rms} \rms \simeq
m_{\rm M}^2/\mcs~~\mbox{with}~~m_{\rm
M}=4\pi\mwr/g^2~~\mbox{and}~~ M_{W^\pm_{\rm
R}}=\sqrt{2}g\vt.\end{equation}
Here $m_{\rm M}$ is the mass of MMs and $\mwr$ the mass of the
gauge bosons. Both are generated by the SSB of $\Glr$ in
\Eref{chain1}. Part of the $\gmcs$ values above can be obtained
for
\beq 1.1\lesssim
M/\YeV\lesssim2.72~~\mbox{and}~~3\lesssim\kp/10^{-4}\lesssim14.\label{rescs1}\eeq
Fixing $\srms=8$, the $M$ values above provide a prediction for
the $\vt$ values
%
\beq 1\leq\vt/\YeV\leq2.86~~\mbox{for}~~0.9\leq M/\YeV\leq2.56\eeq
Therefore, a significant proximity between $\vt$ and $M$ is
requested.

Moreover, the GWs obtained from the CSs have to be consistent with
the upper bound on their abundance $\ogw$ originating from the
advanced \ligo\ third observing run \cite{ligo}
\beq \ogw(f_{\rm LVK})\lesssim 1.7\cdot10^{-8}~~\mbox{for}~~f_{\rm
LVK}=25~{\rm Hz}\label{lvkb}\eeq
which implies $\gmcs\lesssim2\cdot10^{-7}$. The long-lasting
matter domination obtained in our set up due to the $z$
oscillations after the end of FHI -- see \Sref{reh} -- facilitates
the fulfillment of the bound above.

\begin{figure}[!t]\vspace*{-.23in}
\hspace*{-.19in}
\begin{minipage}{8in}
\epsfig{file=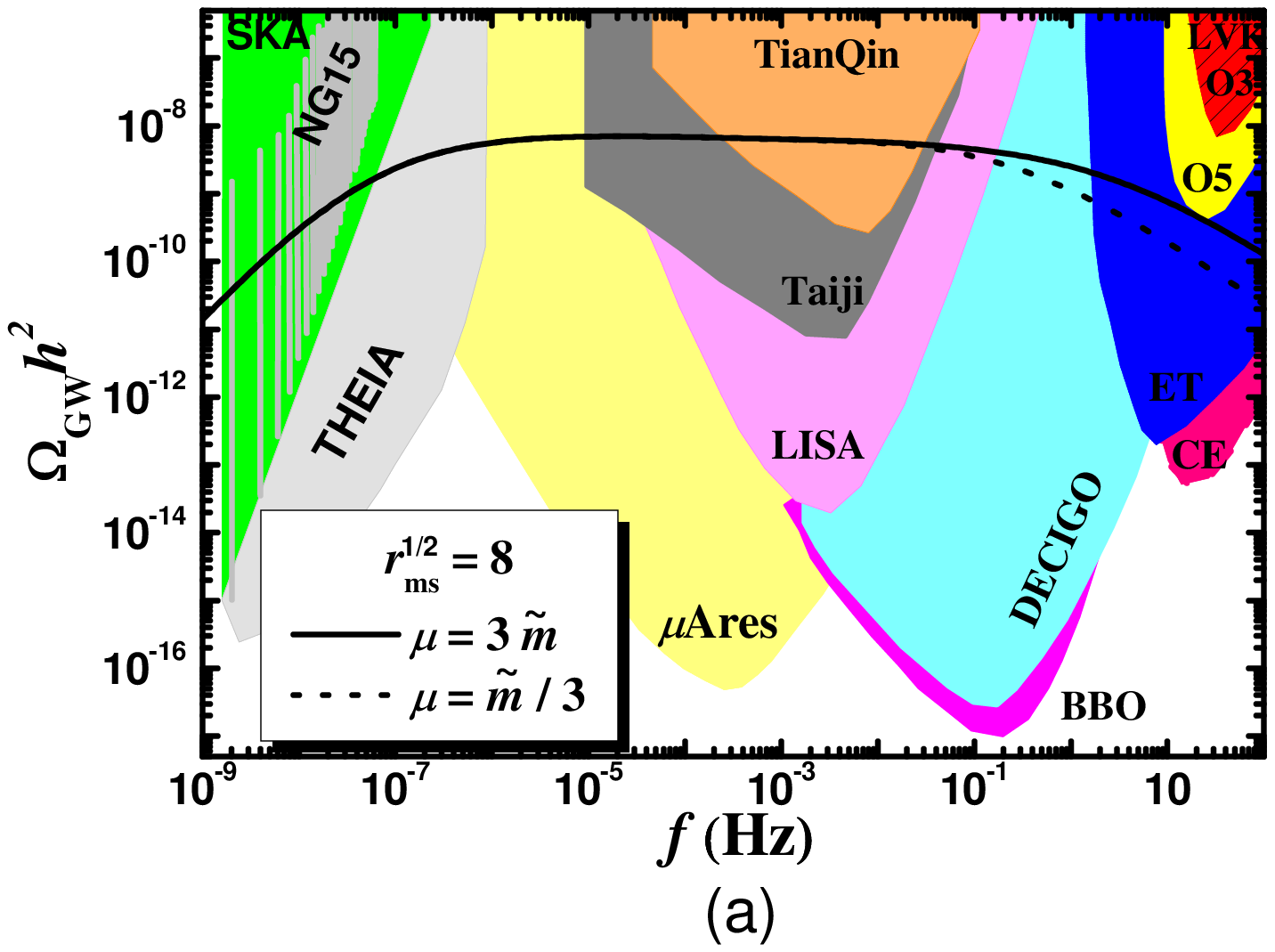,height=3.6in,angle=-90} \hspace*{-1.2cm}
\epsfig{file=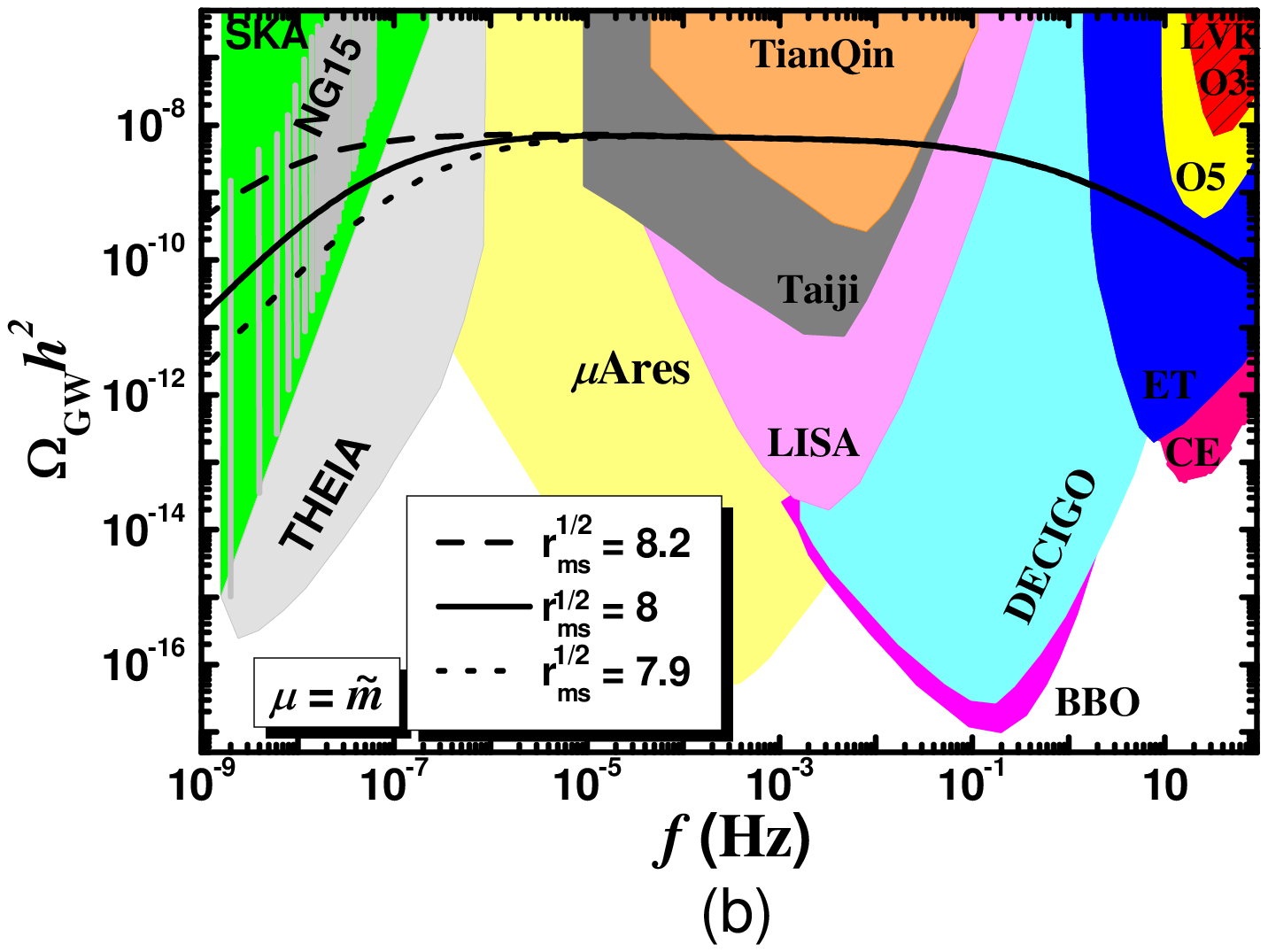,height=3.6in,angle=-90} \hfill
\end{minipage}
\caption{\sl GW spectra from metastable CSs for {\sffamily\ftn
(a)} $\srms=8$  and $\mu=\mss/3$ (dotted line) or $\mu=3\mss$
(solid line) {\sffamily\ftn (b)} $\mu=\mss$ and various $\srms$
indicated in the plot. The remaining inputs are taken from
Table~2. The shaded areas in the background indicate the
sensitivities of various current and future experiments -- see
\cref{blfhi} and references therein.}\label{figw}
\end{figure}



Applying the formulae of \cref{pillado,blfhi} we compute $\ogw$
for the GWs produced from the CS formatted in our setting under
the assumption that these are metastable. Employing the model
parameters of \Tref{tab} -- which yields $\gmcs=7.4\cdot10^{-8}$
-- we obtain the outputs displayed in \Fref{figw}. Namely, in
\sFref{figw}{a} we show $\ogw$ as a function of $f$ for $\srms=8$
and $\mu=\mss/3$ which yields $\Trh=0.4~\GeV$ (dotted line) or
$\mu=3\mss$ which results to $\Trh=3~\GeV$ (solid line) -- see
\Tref{tab}. On the other hand, for the GW spectra depicted in
\sFref{figw}{b} we employ $\mu=\mss$ resulting to $\Trh=1.03~\GeV$
and fix $\srms$ to $7.9$ (dotted line) $8$ (solid line) and $8.2$
(dashed line) -- see \Eref{kai}. In both panels of \Fref{figw} we
see that the derived GW spectra can explain the \nano\ shown with
gray almost vertical lines. We see though that, as $\srms$
increases, the increase of $\ogw$ becomes sharper and provide
better fit to the observations. Also, in both panels the shape of
GW signal suffers a diminishment above a turning frequency $f_{\rm
rh}\sim0.03~\hz$ which enables us to satisfy \Eref{lvkb} more
comfortably than in the case with high reheating. As $\Trh$
decreases, the reduction of $\ogw$ becomes more drastic in
accordance with the findings of \cref{pillado}. The plots also
show examples of sensitivities of possible future observatories
which can test the signals at various $f$ values -- see
\cref{blfhi,lrthi} and references therein.



\section{Conclusions}\label{con}

We analyzed the production of CSs within a $\Glra$-invariant model
which incorporates FHI and SUSY breaking consistently within an
approximate $R$ symmetry. The CSs may be metastable due to
preinflationary GUT genealogy -- see \Eref{chain1}. The model
offers the following interesting, in our opinion, achievements:

\begin{itemize}

\item Observationally acceptable FHI adjusting the tadpole
parameter $\aS$ and the $\Glra$ breaking scale $M$ as shown in
\Eref{res};

\item A prediction of the SUSY-mass scale $\mss$ which turns out
to be of the order of PeV;

\item Generation of the $\mu$ term of MSSM with $|\mu|\sim \mgr$;

\item An interpretation of the DE problem without extensive
tuning;

\item Compatibility of the reheating temperature \Trh\ with BBN;

\item An explanation of the NG15 via the decay of the CSs.

\end{itemize}

Due to the low $\Trh$ our proposal faces difficulties with the
following issues:

\begin{itemize}

\item Baryogenesis. An interesting possibility is to take
advantage from the non-thermal decay of sgoldstino within some
modified version of MSSM -- cf. \cref{allahbau}.

\item Cold-dark-matter abundance. The relic density of the
lightest neutralino although in the $\PeV$ range may be inadequate
to account for the dark-matter abundance. Non-thermal contribution
\cite{kolbdm} from gravitino decay may be welcome for this aim.

\end{itemize}

\paragraph*{\small \bf\scshape Acknowledgments} {\small  I would like to thank R. Maji and Q.
Shafi for useful suggestions, M. Axenides, A. Passias and K.
Tamvakis for interesting questions. }

\def\ijmp#1#2#3{{\sl Int. Jour. Mod. Phys.}
{\bf #1},~#3~(#2)}
\def\plb#1#2#3{{\sl Phys. Lett. B }{\bf #1}, #3 (#2)}
\def\prl#1#2#3{{\sl Phys. Rev. Lett.}
{\bf #1},~#3~(#2)}
\def\rmp#1#2#3{{Rev. Mod. Phys.}
{\bf #1},~#3~(#2)}
\def\prep#1#2#3{{\sl Phys. Rep. }{\bf #1}, #3 (#2)}
\def\prd#1#2#3{{\sl Phys. Rev. D }{\bf #1}, #3 (#2)}
\def\npb#1#2#3{{\sl Nucl. Phys. }{\bf B#1}, #3 (#2)}
\def\npps#1#2#3{{Nucl. Phys. B (Proc. Sup.)}
{\bf #1},~#3~(#2)}
\def\mpl#1#2#3{{Mod. Phys. Lett.}
{\bf #1},~#3~(#2)}
\def\jetp#1#2#3{{JETP Lett. }{\bf #1}, #3 (#2)}
\def\app#1#2#3{{Acta Phys. Polon.}
{\bf #1},~#3~(#2)}
\def\ptp#1#2#3{{Prog. Theor. Phys.}
{\bf #1},~#3~(#2)}
\def\n#1#2#3{{Nature }{\bf #1},~#3~(#2)}
\def\apj#1#2#3{{Astrophys. J.}
{\bf #1},~#3~(#2)}
\def\mnras#1#2#3{{MNRAS }{\bf #1},~#3~(#2)}
\def\grg#1#2#3{{Gen. Rel. Grav.}
{\bf #1},~#3~(#2)}
\def\s#1#2#3{{Science }{\bf #1},~#3~(#2)}
\def\ibid#1#2#3{{\it ibid. }{\bf #1},~#3~(#2)}
\def\cpc#1#2#3{{Comput. Phys. Commun.}
{\bf #1},~#3~(#2)}
\def\astp#1#2#3{{Astropart. Phys.}
{\bf #1},~#3~(#2)}
\def\epjc#1#2#3{{Eur. Phys. J. C}
{\bf #1},~#3~(#2)}
\def\jhep#1#2#3{{\sl J. High Energy Phys.}
{\bf #1}, #3 (#2)}
\newcommand\jcap[3]{{\sl J.\ Cosmol.\ Astropart.\ Phys.\ }{\bf #1}, #3 (#2)}
\newcommand\njp[3]{{\sl New.\ J.\ Phys.\ }{\bf #1}, #3 (#2)}
\def\prdn#1#2#3#4{{\sl Phys. Rev. D }{\bf #1}, no. #4, #3 (#2)}
\def\jcapn#1#2#3#4{{\sl J. Cosmol. Astropart.
Phys. }{\bf #1}, no. #4, #3 (#2)}
\def\epjcn#1#2#3#4{{\sl Eur. Phys. J. C }{\bf #1}, no. #4, #3 (#2)}

\end{document}